\documentstyle[12pt]{article}
\begin{document}
\thispagestyle{empty}
\begin{center}{\Large{On the possibility of primordial torsion detection from magnetic helicity and energy spectra}}
\end{center}
\vspace{1.0cm}
\begin{center}
{\large By L.C. Garcia de Andrade\footnote{Departamento de
F\'{\i}sica Te\'{o}rica - IF - UERJ - Rua S\~{a}o Francisco Xavier
524, Rio de Janeiro, RJ, Maracan\~{a}, CEP:20550.
e-mail:garcia@dft.if.uerj.br}}
\end{center}
\begin{abstract}
The purpose of this paper is twofold: First lending more support to torsion alternative gravity theories to General Relativity and cosmology and torsion detection by showing how they can simply used in the investigation of helicity and magnetic energy spectra of primordial fields by using a strong value of torsion of $10MeV$ previously computed by the author. dynamo mechanism seeds in the Primordial Universe has been a matter of intense investigation lately. The second purpose is to apply these torsion theories in the special random spins gauge where its time component torsion gauge $T^{0}=0$, which implies that spins of the nucleons are polarised orthogonal to QCD domain walls. A modification of circular polarization at $1Mpc$ scale, give rise to left torsion helicity which using a QCD seed $B_{QCD}\sim{10^{16}G}$ yields MF of $B_{-}\sim{10^{30}Gauss}$ which was the value obtained by Enqvist et al using neutrinos and galactic dynamos at QED scales. The $B_{+}$ mode can be shown to constrain torsion to $T\sim{10^{7}Mev}$ at $1pc$ which can be detected at LHC scales. These constraints at 1Mpc reach only $10^{-10}MeV$\end{abstract}
  
Key-words: modified gravity theories, dynamos, Magnetic helicity
\newpage
\section{Introduction}
Earlier Shaw and Lewis \cite{1} investigated the magnetic fields (MF) in the realm of general relativity and addressed the disfavouring galactic  dynamos. Earlier however Gwyn et al \cite{2} have investigated MF from heterotic cosmic string without torsion and obtained galactic dynamo seeds. This seems that with few exceptions like in the Barrow and Tsagas work \cite{3} new physics has to be addeded to obtain such galactic dynamo seeds specially from the early universe. For example Forbes and Zhitnitsky \cite{4} and Kisslinger \cite{5} have succeded with QCD domain walls which represents new physics in the early universe. Also investigations in the very early universe cosmological model have been undertaken by Gasperini \cite{6}. Previous to these attempts Caroll and Field \cite{7} succeded in obtaining these results by investigating the PMF by studying the two modes of magnetic fields on a circular polarisation and from both modes obtained the energy and magnetic helicity spectra bounds to investigate QCD domain walls and obtained the magnetic fields estimates. In this paper we follow the path of their work to extend it to a spacetime endowed by torsion. To apply torsion to this polarised spins of nucleons orthogonal to the QCD wall as in Forbes and Zhitnisky paper, has as motivation Einstein-Cartan theory of gravity as seen by A Trautmann \cite{8} where he used the polarisation of spins by torsion to avoid singularities by holding the gravitational collapse of stars. These ideas can be applied to neutron stars and spin polarisation density is connected to torsion \cite{9}. As an important byproduct of this investigation is that $B_{+}$ modes of MF may give rise to $10^{7}MeV$ strong primordial torsion at 1pc coherent length. The plan of the paper is as follows : In section 2 one applies the dynamo equation in spacetime with torsion derived recently \cite{10} to obtain magnetic field modes solution circular polarised. In section 3 one obtains from these modes the magnetic helicity and magnetic energy spectra and their bounds. In this same section one applies these spectra to domain walls to obtain the magnetic field from QCD. Discussions are left for section 4.
\newpage
\section{Circular Polarised Magnetic modes solutions of Dynamo equation with torsion}
In this section we review briefly the derivation of the dynamo equation with torsion and get solutions where universe presents high resistivity where ${\eta}\sim{10^{28}cm^{2}s^{-1}}$ here $\eta$ is the diffusion or the electric resistivity of the plasma flow. Let us then take now a simple form of solution of the self induction equation with torsion simply given by the minimal circular polarisation solutions for the magnetic field in the form $B_{\pm}$. Before that let us remind the \cite{10} lagrangean interaction from where we have derived the dynamo equation. The lagrangean interaction with parity  violation is given by the action 
\begin{equation}
{S}= \int{d^{4}x(-g)^{\frac{1}{2}}[(-\frac{1}{4}F_{{\mu}{\nu}}F^{\mu\nu}+R_{\mu\nu\rho\sigma}F^{\rho\sigma}F^{\mu\nu}+R_{{\mu}{\nu}{\sigma}{\rho}}{\epsilon}^{{\mu}{\nu}{\sigma}{\rho}}+j^{\mu}A_{\mu})]} \label{1}
\end{equation}
Here $A_{\mu}$ is the electromagnetic field potential, and the e.m field tensor is $F_{\mu\nu}=2{\partial}_{[\mu}A_{\nu]}$ and$R_{{\mu}{\nu}{\alpha}{\beta}}$ is the Riemann-Cartan spacetime curvature endowed with all types of torsion, scalar as used by Bamba et al \cite{11}, vector and axial torsion. From the Euler-Lagrange equation the curvature leads to the dynamo equation
\begin{equation}
{D}_{\mu}F^{\mu\nu}+\frac{1}{m^{2}}D_{\mu}[4{R^{{\mu\nu}}}_{\rho\sigma}F^{\rho\sigma}]=0 \label{2}
\end{equation}
the corrected formula is
\begin{equation}
{\partial}_{\mu}F^{\mu\nu}+T_{\mu}F^{{\mu\nu}}=0
\label{3}
\end{equation}
where we have used $D_{\mu}={\partial}_{\mu}+T_{\mu}$ where $T_{\mu}$ $({\mu}=0,1,2,3)$ , $D_{\mu}$ is the covariant derivative in Riemann-Cartan spacetime. Thus the last expression was obtained taking the torsion field up to first order. In equation (\ref{3}) we have also considered the Bianchi identity
\begin{equation}
D_{\mu}[{R^{{\mu\nu}}}_{\rho\sigma}]=0
\label{4}
\end{equation}
From generalised vector Maxwell equations with torsion one obtains the dynamo equation
\begin{equation}
{\partial}_{t}\textbf{B}-{\nabla}{\times}[\textbf{V}{\times}\textbf{B}]-{\eta}[{\Delta}\textbf{B}-{\nabla}{\times}[\textbf{T}{\times}\textbf{B}]]=0
\label{5}
\end{equation}
when $T$ is in the terrestrial torsion
given in Laemmerzahl \cite{10} given by $T\sim{10^{-17}cm^{-1}}$ or $10^{-31}GeV$ but here we use torsion recently obtained by the author in the early universe \cite{12} which is much stronger of the order of $10MeV=1GeV$. Simple vector analysis yields the simpler equation
\begin{equation}
{\partial}_{t}\textbf{B}={\nabla}{\times}[\textbf{V}{\times}\textbf{B}]+{\eta}[{\Delta}-{\nabla}.\textbf{T}]\textbf{B}
\label{6}
\end{equation}
Note that if we ignore torsion this reduces to the equation of dissipative MHD. By maintaining Caroll Field assumption that the electric field varies very slowly in time in the bulky matter so we can also neglect the velocity $\textbf{V}$ in the dynamo equation so it reduces to
\begin{equation}
{\partial}_{t}\textbf{B}={\eta}[{\Delta}-{\nabla}.\textbf{T}]\textbf{B}
\label{7}
\end{equation}
where space is decomposed $\textbf{B}(\textbf{k})$  considering the Fourier space into helicity modes which is equivalent to use circular polarization, $\textbf{B}(\textbf{k})=B_{+}{\hat{u}}_{+}+B_{-}{\hat{u}}_{-}$; where ${\hat{u}}_{\pm}={{\hat{u}}_{1}}^{+}\pm{i{\hat{u}_{2}}^{-}}$,where ${\hat{u}}_{A}$ (A=1,2) and ${\hat{u}}^{3}=\frac{\textbf{k}}{k}$ form a 3D orthonormal basis. To apply this circular polarization of the magnetic field to dynamo equation one should be write the dynamo equation in Fourier transformation as 
\begin{equation}
{\partial}_{t}\textbf{B}=-{\eta}[k^{2}+i\textbf{k}.\textbf{T}]\textbf{B}
\label{8}
\end{equation}
to simplify even more the circular polarisation we use another auxiliary new magnetic field modes $b_{+}$ and  $b_{-}$ related to the original modes as 
\begin{equation}
{b_{\pm}}={B_{+}{\pm}B_{-}}
\label{9}
\end{equation}
The inverse original modes are
\begin{equation}
{B_{\pm}}=\frac{b_{+}{\pm}b_{-}}{2}
\label{10}
\end{equation}
Then the dynamo equations in the modes b may be expressed as
\begin{equation}
{\partial}_{t}{b_{-}}={\eta}\textbf{k}.\textbf{T}b_{-}
\label{11}
\end{equation}
\begin{equation}
{\partial}_{t}{b_{+}}={\eta}{k}^{2}b_{+}
\label{12}
\end{equation}
These dynamo like equations uncoupled may have easy solutions as
\begin{equation}
{b_{-}}\sim{exp[{\eta}\int{(\textbf{k}\textbf{T})dt}]}
\label{13}
\end{equation}
\begin{equation}
{b_{+}}\sim{exp[-{\eta}{k}^{2}t]}
\label{14}
\end{equation}
Now going back to the original magnetic fields and considering the average integrand of expression (\ref{13}) at the early universe one obtains
\begin{equation}
B_{\pm}(\textbf{k},t)=B_{\pm}(\textbf{k},0)exp[-\eta\int(\textbf{k}.[\textbf{k}\pm\textbf{T}]dt)]
\label{15}
\end{equation}
Note for example that the supression or growth of magnetic field modes $B_{\pm}$ is the sign of the torsion helicity defined by $\textbf{k}\pm\textbf{T}$. For example when the torsion helicity is negative the mode $B_{-}$ is highly suppressed and decays while $B_{+}$ may grow if the following constraint are satisfied
\begin{equation}
{k}\le{T}
\label{16}
\end{equation}
as in the case of dynamo mechanism. This last expression also constraints torsion with the coherence length $\lambda\sim{k^{-1}}$ since for a 1pc coherence length torsion would have a value $T\ge{10^{18}cm^{-1}}\sim{10^{4}GeV}$ of magnetic field since for Other several possibilities exists. In the next section based on the combination of $B_{\pm}$ modes one shall compute the magnetic energy and helicities densities. Before to close this section we shall be concerned with the substitution in above equation by the inverse case where we consider the gauge where $T^{0}$ is the only component that survives as in cosmological neutrino sea where spins are not polarised \cite{9}. Taking $T^{0}=\dot{\phi}$ where $\phi$ is a torsion pseudoscalar potential one obtains in the integrand of the above expressions a result similar to Carroll Field one as given by
\begin{equation}
B_{\pm}(\textbf{k},t)=B_{\pm}(\textbf{k},0)exp[-\eta\int{{k}.[{k}\mp\dot{\phi}]dt}]
\label{17}
\end{equation}
This expression is exactly the same result of Carrol and Field which hands support to our theory. In their case when $\dot{\phi}$ vanishes $B_{-}$ modes also undergoes Ohmic decay.
\newpage

\section{Energy and Magnetic helicity Spectra from Primordial Torsion}
Magnetic field helicity is a very important issue in what concerns inverse cascades and going from small to large scale magnetic fields in the universe. In the present section we showed that the magnetic helicity has a clear contribution from torsion helicity. Computation of energy and helicity spectra are given by 
\begin{equation}
{{e}^{M}}_{k}=4{\pi}{k}^{2}({|B_{+}|}^{2}+{|B_{-}|}^{2})
\label{18}
\end{equation}
\begin{equation}
{{h}^{M}}_{k}=8{\pi}{k}({|B_{+}|}^{2}-{|B_{-}|}^{2})
\label{19}
\end{equation}
From the expressions of the previous section we are able to compute the energy and helicity spectra in the torsion case as
\begin{equation}
{{e}^{M}}_{k}\ge{8{\pi}{k}^{2}}|B_{seed}|^{2}e^{-{\eta}k^{2}t}
\label{20}
\end{equation}
\begin{equation}
{{h}^{M}}_{k}\le{6{\pi}{k}^{2}|B_{seed}|^{2}e^{-{\eta}\textbf{k}.\textbf{T}t}}
\label{21}
\end{equation}
where to obtain these bounds we made use of the well-known Schwarz inequalities. Now let us note that the energy density lower bound means that the magnetic energy is not suppressed as time evolves, while in the new torsion case upper bound the magnetic helicity growth or decay depends solely upon the sign of torsion helicity. If the torsion helicity is positive the helicity of MF decays and is highly supressed while the case when it is posititive renders room for the magnetic helicity to grow in a spacetime with torsion. To close this section we apply these results to QCD walls to compute the value of the magnetic field there in terms of the strong torsion above. This be simply obtained by taking the data of reference \cite{5}. But to this aim let us first compute the magnetic modes in the approximation
\begin{equation}
{B}_{\pm}=\frac{1}{2}B_{seed}[(1-\eta{k}^{2}t)\pm(1-\eta\textbf{T}.\textbf{k})t]
\label{22}
\end{equation}
Note that $B{-}$ MF mode reduces a very simple form to compute the $B_{QCD}$ field. This $B_{-}$ mode reduces to
\begin{equation}
{B}_{-}=\frac{1}{2}B_{seed}[\eta[-{k}^{2}+\textbf{T}.\textbf{k}]t]
\label{23}
\end{equation}From this last expression one easily obtains that in $B_{-}$ mode only torsion helicity contribution sutvive as universe expands then\begin{equation}
{B}_{-}\approx{\frac{1}{2}B_{seed}[\eta[\textbf{T}.\textbf{k}]t]}
\label{24}
\end{equation}
and again the $B_{-}$ mode cannot be suppressed if the torsion helicity is positive. Thus this mode contributes to a dynamo action. From this last expression one easily obtain just from torsion contribution 
\begin{equation}
{B}_{-}\approx{10^{11}Gauss}
\label{25}
\end{equation}
This result was obtained by noticing that at QCD domain wall 
\begin{equation}
{\tau}\sim{\frac{1}{{\Lambda}_{QCD}}}\sim{\frac{1}{0.15GeV}}\label{26}
\end{equation}
so taking into consideration that $T\approx{k^{-1}}$ and that $1Gev\sim{10^{14}cm^{-1}}$ from the expression
\begin{equation}
B_{-}\sim{B_{QCD}\eta{kT}{\tau}}\sim{\frac{B_{QCD}{\eta}kT}{{\Lambda}_{QCD}}}\label{27}
\end{equation}
where we take into account that at 1Mpc gives with a seed field of $B_{CQD}\sim{10^{16}G}$. So the left handed mode of magnetic field from QCD seeds is 
\begin{equation}
{B_{-}}^{PMF}\sim{10^{30}Gauss}\label{28}
\end{equation}
which is closed to the value obtained by Enqvist et al \cite{14} with galactic dynamos and neutrinos. 
\section{Conclusions}
The dynamo mechanism seeds in the Primordial Universe has been a matter of intense investigation lately. Here we solved the dynamo equation recently derived by the author, with torsion vector in first approximation using the torsion gauge $T^{0}=0$. The gravitational physics behind this gauge is that the spins are polarised as happens in QCD domain walls. A modification of circular polarisation modes of  Carroll and Field to investigate Primordial Magnetic Fields (PMF) is done in this paper to accomodate torsion in the spacetime of circular polarised magnetic field modes. From these solutions one obtains the magnetic helicity and magnetic energy spectra. One obtains that the energy spectrum does not depend explicity on torsion while the magnetic helicity spectra does. This is a new result since magnetic helicity spectra bounds are shown to depend on primordial torsion helicity. On the other hand we showed that when spins are randomically distributed as in a neutrino cosmological sea, since only $T^{0}$ component of torsion survives we can replaced it in the dynamo equation and it is similar to Carroll and Field pseudo-scalar fields when $T^{0}$ is the time derivative of a pseudo-scalar field like a torsion potential. In the new case where the spins are polarised by the torsion vector and magnetic field solutions only some modes of magnetic field grow while others decay depending on magnetic helicity sign. In the case of torsion vectors this can be applied to QCD domain walls to compute the order of estimate of magnetic fields. Magnetic field mode not suppressed where torsion dominates and with a QCD field of $B_{QCD}\sim{10^{16}G}$ at 1Mpc one obtains a magnitude of $B_{-}(PMF)\sim{10^{30}Gauss}$ which was the magnitude obtained by Enqvist et al \cite{14} using neutrinos. Future investigations of to check if torsion, even in strong gravity case can give rise to the so-called torsion dynamos, in very exceptional situations where $M_{GUT} \sim{10^{17}MeV}$ \cite{15} when torsion in supergravity non-standard model for particle physics would be useful. Torsion detection which is an old theme in gravitational physics \cite{16} can be obtained here from MF right hand mode which yields a constraints on spacetime-torsion of the order of $10^{7}MeV$.
As was easily be showned above magnetic helicity instability may driven dynamos while other modes are highly supressed. Remains then the possibility of explointing torsion in non-standard quantum gravity model of supergravity to investigate its deep relation with early universe. 
\section{Acknowledgements}
We would like to express my gratitude to Tina Kahniashvili and Ariel Zhitnitsky for helpful discussions on the subject of this paper specially to Ariel for some discussions and correspondences on torsion effects on QCD domain walls. Financial support University of State of Rio de Janeiro (UERJ) is grateful
acknowledged.

\end{document}